\documentclass{article}
\usepackage{spconf,amsmath,epsfig}
\usepackage{multirow}
\usepackage{subfigure}

\begin{document}\sloppy

\def\x{{\mathbf x}}
\def\L{{\cal L}}

\title{Robust Contrast Enhancement Forensics Using Pixel and Histogram Domain CNNs}
%
\name{Pengpeng Yang, Rongrong Ni*, Yao Zhao, Gang Cao, Wei Zhao}
\address{Institute of Information Science, Beijing Jiaotong University, Beijing, China \\ Beijing Key Laboratory of Advanced Information Science and Network Technology \\Communication University of China, Beijing, China}

\maketitle

\begin{abstract}
Contrast enhancement (CE) forensics has always been of concern to image forensics community. It can provide an effective tool for recovering image history and identifying tampered images. Although several CE forensic algorithms have been proposed, their robustness against some processing is still unsatisfactory, such as JPEG compression and anti-forensic attacks. In order to attenuate such deficiency, in this paper we first present a discriminability analysis of CE forensics in pixel and gray level histogram domains. Then, in such two domains, two end-to-end methods based on convolutional neural networks (P-CNN, H-CNN) are proposed to achieve robust CE forensics against pre-JPEG compression and anti-forensics attacks. Experimental results show that the proposed methods achieve much better performance than the state-of-the-art schemes for CE detection in the case of no other operation and comparable performance when pre-JPEG compression and anti-foresics attacks is used. 
\end{abstract}
\section{Introduction}
\label{sec:intro}

As a simple yet efficient image processing operation, CE is typically used by malicious image attackers to eliminate inconsistent brightness for generating a visually imperceptible tampered images. CE detection algorithms play an important role in decision analysis for authenticity and integrity of digital images. Although some schemes have been proposed to detect contrast-enhanced images, the performance of such techniques is limited in the cases of pre-JPEG compression and anti-forensic attacks. Therefore, it is critical to develop robust and effective CE forensics algorithms.

With the efforts of researches in the past decade, a number of schemes [1-9] have been proposed to discriminate the contrast-enhanced images in uncompressed format. Stamm \emph{et al.} [1,2,3] found that contrast enhancement would introduce the peaks and gaps into the image's gray level histogram, which led to the specific high values in high-frequency components. Lin \emph{et al.} [6,7] revealed that contrast enhancement would disturb the inter-channel correlation left by color image interpolation and measured such correlation to distinguish the original and enhanced images. Futhermore, in order to recover the image processing history, the algorithms [10-13] of estimating parameters for constrast-enhanced images are developed.

Despite good performance of the above algorithms, their robustness is unsatisfactory in some cases, such as the CE of JPEG images (pre-JPEG compression) and the occurrence of anti-forensic attacks [14-19]. The reason lies in that the fingerprint left by CE operation would be destroyed. Based on such a phenomenon, some researchers attempt to propose more robust CE forensic algorithms, which contains two major branches: overcoming pre-JPEG compression [4] and defensing anti-forensic attack [9]. Unfortunately, such methods can not address well both pre-JPEG compression and anti-forensic attacks. And to date there are no satisfactory solutions.



In this paper, we propose two robust CE detection algorithms based on convolutional neural networks (CNNs) to resist not only pre-JPEG compression but also anti-CE attacks. Firstly, discriminability analysis of CE forensics in pixel and histogram domains is presented. Then, inspired by the excellent performance of deep learning based techniques in various fields, we explore two types of CNNs architectures for CE forensics: pixel-domain CNNs (P-CNN) and histogram-domain CNNs (H-CNN). Especially for P-CNN, high-pass filter is used to reduce the affect of image contents and keep the data distribution balance cooperating with batch normalization [21]. Additionally, the width of architecture is experimentally designed to learn better feature representation for CE forensics. Besides, as a lower dimensional yet effective feature, the histogram with 256 dimensions is fed into CNNs for constructing H-CNN. Experimental results show that our proposed methods outperform the state-of-the-arts schemes in the case of uncompression and comparable performance in the cases of pre-JPEG compression, anti-forensics attack, and CE level variation.

\section{Proposed Robust Algorithm for detecting Contrast Enhancement Images}

  The existing algorithms are not robust against pre-JPEG compression, anti-forensics attack and CE level variation. In this paper, two deep learning-based algorithms, data-driven framework, are proposed to detect contrast-enhanced images by auto-learning effective features from database: P-CNN and H-CNN. Specifically, their architectures are as shown in Fig.1.

  \begin{figure*}[t]
  	\centering
  	\subfigure{\begin{minipage}{11cm}\includegraphics[width=11cm,height=5.5cm]{{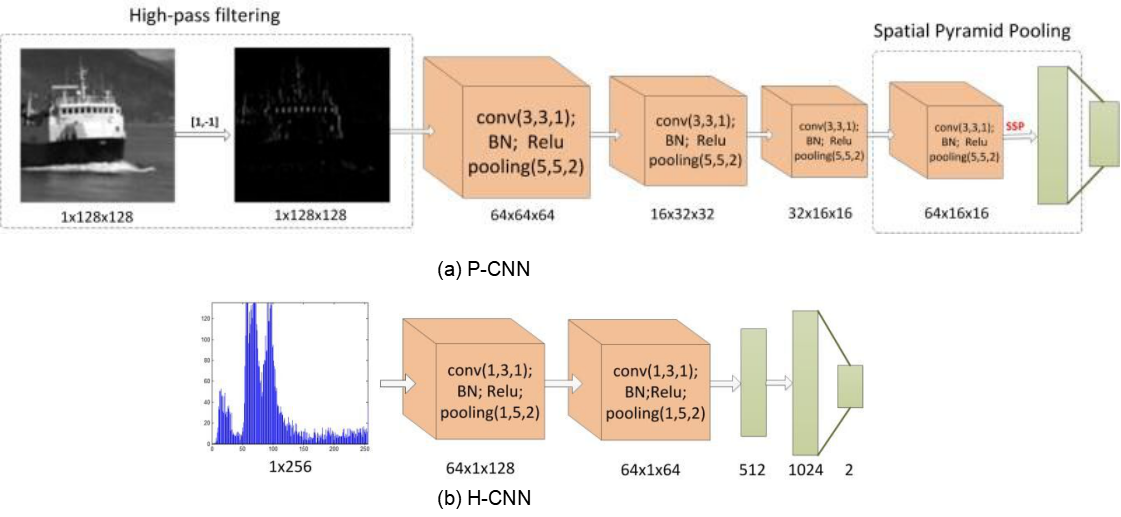}}\end{minipage}}
  	\hspace{0.0001cm}
  	
  	\caption{The architecture of proposed pixel-domain (a) and histogram-domain (b) convolutional neural networks.}
  \end{figure*}

\subsection{Pixel-Domain Convolutional Neural Networks}

As a common way of contrast enhancement, gamma correction can be found in many image-editing tools. In this paper, we mainly focus on the detection of gamma correlation, which is typically defined as,
\begin{equation}\label{key}
Y=[255(X/255)^\gamma ]\approx 255(T^\gamma)  
\end{equation}
where $X$ denotes an input and $Y$ represents the mapped value, $T=(X/255) \epsilon [0,1]$. In order to simplify the discussion, the mapped value, $Y$, is normalized: 
\begin{equation}\label{key}
Z=Y/255\approx T^\gamma  
\end{equation}
where $Z\epsilon [0,1]$. As well known, gamma correction would lead to the nonlinear changes in pixel domain and introduce the peak/gap bins into histogram domain [1-4]. A number of handcrafted features are designed based on such phenomenons. 

In pixel domain, the difference between the original and enhanced images can be computed as follows, and the absolute value of difference is considered. 
\begin{equation}
\left\{
\begin{array}{lr}
D = Y-X=255(Z - T) \approx 255(T^\gamma-T), \gamma <1  \\ 
\\
D = X-Y=255(T - Z) \approx 255(T-T^\gamma), \gamma >1 
\end{array}
\right.
\end{equation}

It can be seen from (3) that the discriminability in pixel domain is related with pixel value (image contents), $T$, and parameter of gamma correction, $\gamma$. In order to describe such discriminability, the maximum of difference denoted by $D_{max}$ is considered. $D_{max}$ is obtained when partial derivative of $Z$ with respect to $T$ is equal to $1$.
\begin{equation}\label{key}
T_{D_{max}} = T_{\frac{\partial Z}{\partial T}=1}=(\frac{1}{\gamma })^{(\frac{1}{\gamma -1})} 
\end{equation}

\begin{equation}\label{key}
D_{max} = 
\left\{
\begin{array}{lr}
255[(\frac{1}{\gamma })^{\frac{\gamma }{\gamma -1}} - (\frac{1}{\gamma })^{\frac{1}{\gamma -1}}],\gamma <1  \\ 
\\
255[(\frac{1}{\gamma })^{\frac{1}{\gamma -1}} - (\frac{1}{\gamma })^{\frac{\gamma }{\gamma -1}}],\gamma >1
\end{array}
\right.
\end{equation}

The curve of function of $D_{max}/255$ on $\gamma$ is shown in Fig.2. For the purposes of understanding, four groups of parameters are chosen in the following discussion: $\gamma=\{0.6,0.8,1.2,1.4\}$. It is easy to find that $D_{max_A}$($\gamma=0.6$)$=47.4045$ $>$ $D_{max_D}$($\gamma=1.4$)$=31.416$ $>$$D_{max_B}$($\gamma=0.8$)$=20.8896$ $>$$D_{max_C}$($\gamma=1.2$)$=17.0799$. 

\begin{figure}[!h]
	\centering
	\includegraphics[width=4cm,height=3cm]{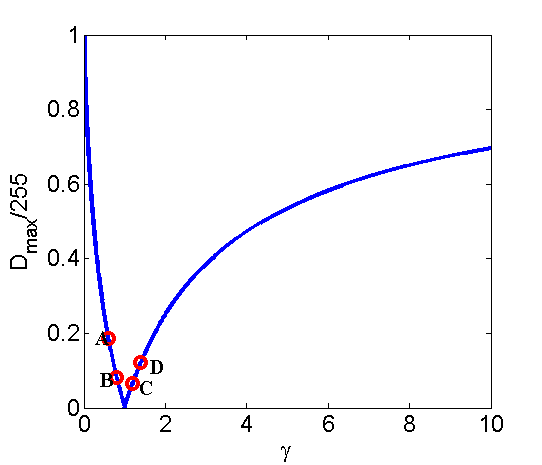}
	\caption{The curve of function of $D_{max}/255$ on $\gamma$ and $A, B, C, D$ are $\gamma$=0.6, 0.8, 1.2, 1.4, respectively.}
\end{figure}

Fortunately, in spite of the changes of discriminability in pixel domain, the difference in pixel domain could be learned by deep learning-based method. Inspired by it, the P-CNN is proposed to detect enhanced image. The design of P-CNN is as follows.

Firstly, the high-pass filter is added into the front-end of architecture to eliminate the interfere of image content. Another advantage of using high-pass filter could be that it accelerates training by cooperating with batch normalization. Because that the histogram of high-pass filtered images approximately follows the generalized Gaussian distribution, which is similar to batch normalization [21]. In particular, we experimentally find that the filter of the first-order difference along horizontal direction has better performance. 

\begin{equation}\label{key}
I_{1} = H*I
\end{equation}
where $H=[1,-1]$, $I$ is the input image, $I_{1}$ is the output of the first layer, '*' represents the convolution operator.

Next, high-pass filtering layer are followed by four traditional convolutional layers. For each layer, there are four types of operations: convolution, batch normalization, ReLU and average pooling. The feature maps for each layer are 64, 16, 32, 128, respectively. The kernel size for convolutional and pooling operation is 3x3 with 1 stride, 5x5 with 2 strides. It should be pointed out that: 1) we experimentally find that the numbers of feature map for first convolutional layer is important for CE detection and it has better performance when the feature maps is 64. In other words, low-level feature would be more helpful; 2) instead of average pooling, the spatial pyramid pooling layer [20] is used in last convolutional layer to fuse multi-scale features. The convolutional layer is calculated as

\begin{equation}\label{key}
I_{i}=
\left\{
\begin{array}{lr}
P(R(F(W_{i}*I_{i-1}+B_{i}))), i\epsilon (2,3,4)\\ 
S(R(F(W_{i}*I_{i-1}+B_{i}))), i=5
\end{array}
\right.
\end{equation}
where $F, R, P ,S$ represents the batch normalization, ReLU, average pooling, and spatial pyramid pooling, respectively. For spatial pyramid pooling, three scales are chosen and lead to 2688 dimensional output.

In the end, the fully connected layer and softmax is followed by a multinomial logistic loss. The loss function is defined as,
\begin{equation}\label{key}
Loss=-log(\frac{e^{W^{j}I_{5}+B^{j}}}{\sum_{j=1}^{n}e^{W^{j}I_{5}+B^{j}}})
\end{equation}
where $n$ is the number of classes and $j$ denotes the true label. In our experimental setup, Mini-batch Stochastic Gradient Descent is applied and the batch size is set as 120. The learning rate is initialized as 0.001, and scheduled to decrease 10\% for every 10000 iterations. The max iterations is 100000. The momentum and weight\_decay are fixed to 0.9 and 0.0005, respectively.

\subsection{Histogram-Domain Convolutional Neural Networks}

According to the report [4], the handcrafted feature based on histogram is also vulnerable. The peak and gap feature is easily destroyed by pre-JPEG compression and anti-forensic attacks. In order to detect the CE of JPEG compressed images, Cao \emph{et al.} only used the numbers of gap bins as features. However, its performance for different gamma parameters is unstable and it does not work for anti-forensics attack, which could be caused by the unsteadiness of gap bins. 

The reason why gamma correction could cause gap bins is that a strait range of values is projected to the wide one. For example, the values in the range $\rm \left [ 0,T_{D_{max}}^{} \right ], \gamma <1$ will be changed to the range of $\rm \left [ 0,T_{D_{max}}^{\gamma} \right ]$. Therefore, the probability of gap bins (zero bins) should be proportional to the ratio of wide range of values and corresponding strait range,

\begin{figure*}[!t]
	\centering
	\subfigure{\begin{minipage}{3.3cm}\includegraphics[width=3.3cm,height=3cm]{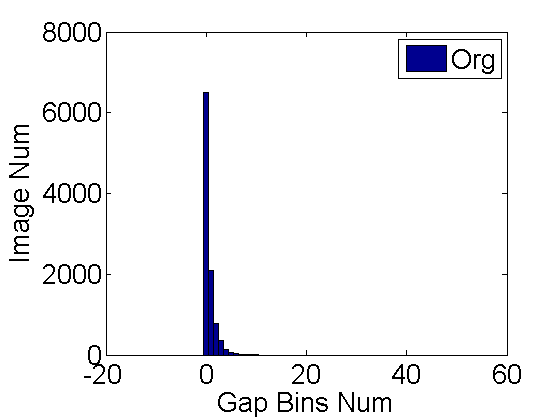}\end{minipage}}
	\hspace{0.0001cm}  
	\subfigure{\begin{minipage}{3.3cm}\includegraphics[width=3.3cm,height=3cm]{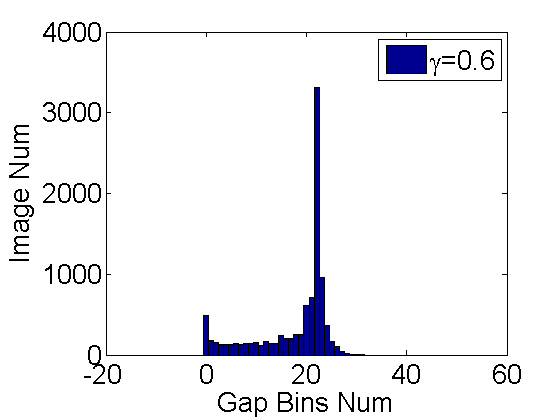}\end{minipage}}  
	\hspace{0.0001cm}
	\subfigure{\begin{minipage}{3.3cm}\includegraphics[width=3.3cm,height=3cm]{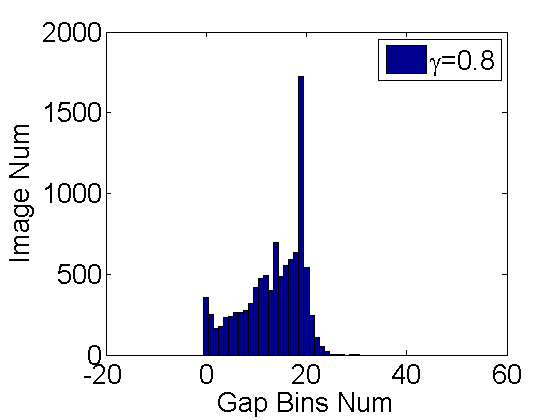}\end{minipage}}  
	\hspace{0.0001cm}
	\subfigure{\begin{minipage}{3.3cm}\includegraphics[width=3.3cm,height=3cm]{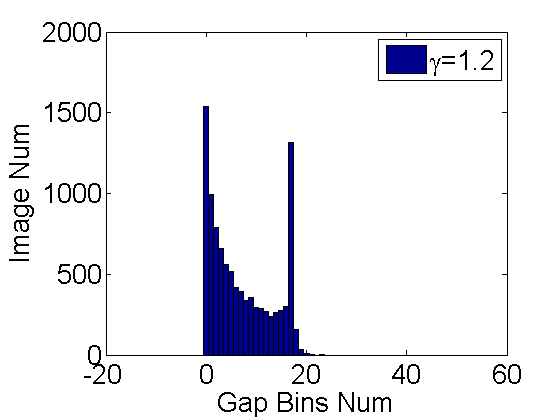}\end{minipage}}   
	\hspace{0.0001cm}
	\subfigure{\begin{minipage}{3.3cm}\includegraphics[width=3.3cm,height=3cm]{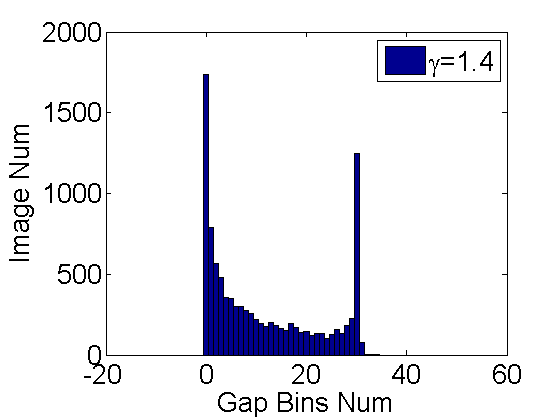}\end{minipage}}   
	\caption{ The statistical distribution of gap bins for original and contrast-enhanced images with different parameters. The images are from BOSSBase dataset and centrally cropped into 128x128 patches.}
\end{figure*}

\begin{equation}\label{key}
\begin{array}{lr}
P_{zero\_bin}\propto G\left ( r \right ) = 
\left\{
\begin{array}{lr}
\frac{T_{D_{max}}^{\gamma }-T_{D_{max}}}{T_{D_{max}}} = \frac{1}{\gamma }-1,\gamma <1\\ 
\\
\frac{T_{D_{max}}-T_{D_{max}}^{\gamma }}{1-T_{D_{max}}} = \frac{(\frac{1}{\gamma })^{\frac{1}{\gamma -1}} - (\frac{1}{\gamma })^{\frac{\gamma }{\gamma -1}}}{1-(\frac{1}{\gamma })^{\frac{1}{\gamma -1}}},\gamma >1
\end{array}
\right.
\end{array}
\end{equation}

It can be found that  $G(0.6)>G(0.8)>G(1.4)>G(1.2)$  
, which means that the numbers of gap bins is ranged among CE parameters. The statistical distribution of gap bins for the original and enhanced images with $\gamma=0.6,0.8,1.2,1.4$ is shown in Fig. 3. As can be seen that the numbers of gap bins for $\gamma=0.6, 0.8$ are larger than $\gamma=1.2,1.4$ and the overlapping parts with original images for $\gamma=0.6, 0.8$  are less than $\gamma=1.2,1.4$, which is consistent with the result of our theoretical analysis. Despite the instability of peak/gap bins, we believe that the effective feature could be auto-learned from histogram domain by using data-driven algorithm. Instead of designing features, the histogram-domain convolutional neural networks is constructed to achieve end-to-end self-learning detection. The H-CNN is proposed to self-learn better feature directly from histogram domain. In addition, as an input with low and fixed dimension, the histogram is suitable for convolutional neural networks. The architecture of H-CNN is shown in Fig 1 (b). Its input is the histogram of the image, namely a vector with 1x256 dimensions. Then, such an input layer is followed by two convolutional and three fully connected layers. The feature maps are 64, 64, 512, 1024, 2, respectively. Lastly, the softmax layer followed by a multinomial logistic loss is added to classify original and enhanced images. The parameters of convolutional layers and hyper-parameters are the same as the P-CNN.



\section{Experimental Results}
In order to verify the validity of proposed methods, four groups of experiments are conducted: ORG VS P\--CE, JPEG\--ORG VS JPEG\--CE, ORG VS Anti\--CE, and JPEG\--ORG VS JPEG\--CE\--Anti\--CE,  where ORG is original images in uncompressed format, JPEG\--ORG represents original images in JPEG format, P\--CE and JPEG\--CE denote enhanced versions of ORG and JPEG\--ORG, respectively, and Anti\--CE and JPEG\--CE\--Anti\--CE represent enhanced images with anti-forensics attack for P\--CE, JPEG\--CE, respectively. The BOSSBase [22] with 10000 images is chosen to construct the dataset. Firstly, the images are centrally cropped into 128x128 pixel patches as ORG. Then, JPEG compression with $Q=70, 50$ is carried out for ORG to build JPEG\--ORG. Next, gamma correction with $\gamma=\{0.6, 0.8, 1.2, 1.4\}$ is implemented on ORG, JPEG\--ORG to constitute P\--CE and JPEG\--CE. In the end, Anti\--CE is produced by anti-forensics attacks [12,14] on P\--CE. It should be noted that the reasons for our choice of pixe patch size are that 1) the detection for the images with lower resolution is much harder than higher resolution image; 2) 128x128 is a suitable size for tamper locating based on CE forensics; 3) our hardware configuration is limited. For each experiment, the training data, validation and testing data is 8000, 2000, 10000, respectively. The experiments about the proposed schemes are conducted on one GPU (NVIDIA TITAN X) with an open source framework of deep learning: Caffe [23].

\subsection{Contrast Enhancement Detection For Contrast-Enhanced Images}
The result for contrast-enhanced images in uncompressed format, is as shown in Table I. P-CNN is pixel-domain convolutional neural networks and H-CNN is histogram-domain convolutional neural networks. As seen from the Table 1, for Cao's method, the detection accuracy for $\gamma=\{0.6, 0.8\}$ is much higher than one for $\gamma=\{1.2, 1.4\}$. The reason is that gap feature is unstable among CE parameters, which is consistent with our analysis in Section II. 

Inspired by transfer learning techniques [24], we further improve performance of P-CNN by finetuning the model for $\gamma=\{0.8, 1.2, 1.4\}$ from the model for $\gamma=0.6$. P-CNN\--FT achieve better performance than De Rosa's and Cao's methods and H-CNN have much better performance than the state-of-the-art schemes. It should be noted that the performance of H-CNN is better and more stable than the others. Such results demonstrated that the histogram domain feature should be effective for CE detection.

\begin{table}[!h]
	\renewcommand\arraystretch{1.1}
	\centering 
	\caption{The detection accuracy for contrast-enhanced images.}
	\label{my-label}
	\begin{tabular}{p{1.5cm}|p{1.3cm}p{1.3cm}p{1.3cm}p{1.3cm}}
		\hline
		Method & $\gamma=0.6$    &  $\gamma=0.8$  & $\gamma=1.2$ & $\gamma=1.4$    \\ \hline 
		De Rosa[9] & 94.02\%    &  84.85\%  & 78.37\% & 74.12\%  \\ 
		Cao[4] & 93.89\%    &  93.90\%  & 80.26\% & 81.40\%  \\ 
		Li[5] & 93.63\%    &  89.48\%  & 90.76\% & 93.44\%  \\ 
		P-CNN   & 94.70\%    &  89.00\%  & 78.00\% & 86.00\%  \\ 
		P-CNN \_FT  & 94.70\%    &  90.00\%  & 82.45\% & 88.00\%  \\ 
		H-CNN & \textbf{99.48\%}    &  \textbf{99.45\%}  & \textbf{99.40\%} & \textbf{99.07\%}  \\ \hline
	\end{tabular}
\end{table}

\subsection{Robustness Against Pre-JPEG Compressed and Anti-Forensic Attacked Contrast-Enhanced Images}
The performance of different methods for pre-JPEG compressed images with $Q=\{50, 70\}$ and anti-forensics attacked images are shown in Table 2,3,4 and 5. It can be seen from Table II that P-CNN and H-CNN have much higher detection accuracy than De Rosa's and Cao's methods and comparable performance with Li's method. Besides, there is an interesting phenomenon that the performance of P-CNN has a significant improvement compared to P\--CE detection. The reason may be attributed to that JPEG compression weakens the signal components in high frequence and the difference between original and enhanced images after JPEG compressing would be highlighted.

%

\begin{table}[!t]
	\renewcommand\arraystretch{1.1}
	\centering 
	\caption{The detection accuracy for JPEG compressed image with different QFs.}
	\label{my-label}
	\begin{tabular}{p{1cm}|c|p{1cm}p{1cm}p{1cm}p{1cm}p{1cm}}
		\hline
		QF                & Method    &$\gamma=0.6$  & $\gamma=0.8$  & $\gamma=1.2$  & $\gamma=1.4$  \\ \hline
		\multirow{4}{*}{50}    &De Rosa[9]      & 81.50\%  & 79.69\%  & 75.16\%  & 72.70\%       \\ 
		& Cao[4]   & 93.96\%  & 93.75\%  & 80.36\%  & 81.57\%    \\  
		& Li[5]   & 99.11\%  & 98.59\%  & 97.75\%  & 98.43\%    \\  
		& P-CNN      & 98.20\%  & 98.25\%  & 96.70\%  & 97.30\%  \\  
		& H-CNN      & \textbf{99.90\%}  & \textbf{99.80\%}  & \textbf{99.50\%}  & \textbf{99.78\%}     \\   \hline
		
		\multirow{4}{*}{70}    &De Rosa[9]    & 83.99\%  & 82.27\%  & 77.47\%  & 72.95\%      \\  
		& Cao[4]   & 94.06\%  & 93.77\%  & 80.55\%  & 81.56\%    \\  
		& Li[5]   & 98.54\%  & 97.42\%  & 96.22\%  & \textbf{97.79\%}    \\  
		& P-CNN      & 98.60\%  & 97.00\%  & 95.70\%  & 96.50\%  \\  
		& H-CNN      & \textbf{98.86\%}  & \textbf{99.03\%}  & \textbf{98.27\%}  & 97.68\%     \\   \hline 
	\end{tabular}
\end{table}

For anti-forensic attacks, Cao's method does not work and there is a degradation in performance of H-CNN, especially, when anti-forensic method [12] is applied. Because that the anti-forensic attacks would conceal the peak/gap feature in histogram domain. In addition, the anti-forensics attacks based on histogram maybe has no or slight effect on pixel domain. Therefore, the P-CNN has best performance in this case. While the pre-compression and anti-forensic attack are put into together, as shown in Table 5, the proposed method have comparable with Li' scheme.

In conclusion, De Rosa's method is not robust for pre-JPEG compression and anti-forensics attack and Cao's method is vulnerable for anti-forenisic attack. Furthermore, such prior algorithms are unstable in different gamma levels. Although Li's method is better than previous works in the case of pre-JPEG compression and anti-forensic attack, its performance is unsatisfactory when no other operation is used. Comparing with the above schemes, the proposed P-CNN and H-CNN, achieve good robustness against pre-JPEG compression, anti-forenic attack, and CE level variation and H-CNN achieve much better performance in the case of no other operation.

\begin{table}[!t]
	\renewcommand\arraystretch{1.1}
	\centering 
	\caption{The detection accuracy in the case of anti-forensics attack [14].}
	\label{my-label}
	\begin{tabular}{p{1.5cm}|p{1.2cm}p{1.2cm}p{1.2cm}p{1.2cm}}
		\hline
		Method & $\gamma=0.6$    &  $\gamma=0.8$  & $\gamma=1.2$ & $\gamma=1.4$     \\ \hline
		De Rosa[9] & 69.85\%    &  66.03\%  & 62.29\% & 64.42\%  \\ 
		Cao[4]  & \------  & \------    & \------ & \------  \\ 
		Li[5]  &\textbf{99.57\%}  & \textbf{99.38\%} & \textbf{99.33\%} &\textbf{99.51\%}  \\
		P-CNN  & 98.6\%    &  98.5\%  & 97.8\% & 98\%  \\ 
		H-CNN & 98.82\%    &  97.59\%  & 97.57\% & 97.09\%     \\ \hline
	\end{tabular}
\end{table}

\begin{table}[!t]
	\renewcommand\arraystretch{1.1}
	\centering 
	\caption{The detection accuracy in the case of anti-forensics attack [12].}
	\label{my-label}
	\begin{tabular}{p{1.5cm}|p{1.2cm}p{1.2cm}p{1.2cm}p{1.2cm}}
		\hline
		Method & $\gamma=0.6$    &  $\gamma=0.8$  & $\gamma=1.2$ & $\gamma=1.4$     \\ \hline
		De Rosa[9] & 61.674\%    &  58.826\%  & 55.320\% & 59.329\%  \\ 
		Cao[4]  & \------  & \------    & \------ & \------  \\ 
		Li[5]  &96.3\%  & 95.54\% & 95.72\% &96.55\%  \\
		P-CNN  & \textbf{97.9\%}    &  \textbf{96\%}  & \textbf{96.5\%} & \textbf{96.55\%}  \\ 
		H-CNN & 88.77\%    &  73.65\%  & 74.85\% & 78.42\%     \\ \hline
	\end{tabular}
\end{table}

\begin{table}[!t]
	\renewcommand\arraystretch{1.1}
	\centering 
	\caption{The detection accuracy for JPEG compressed image with different QFs and anti-forensics attack [14].}
	\label{my-label}
	\begin{tabular}{p{1cm}|c|p{1cm}p{1cm}p{1cm}p{1cm}p{1cm}}
		\hline
		QF                & Method    &$\gamma=0.6$  & $\gamma=0.8$  & $\gamma=1.2$  & $\gamma=1.4$  \\ \hline
		\multirow{4}{*}{50}    &De Rosa[9]      & 70.26\%  & 67.85\%  & 65.38\%  & 66.52\%       \\ 
		& Cao[4]    & \------  & \------    & \------ & \------  \\ 
		& Li[5]   & \textbf{99.9\%}  & \textbf{99.9\%}  & \textbf{99.9\%}  & \textbf{99.9\%}    \\  
		& P-CNN      & \textbf{99.9\%}  & \textbf{99.9\%}  & \textbf{99.9\%}  & \textbf{99.9\%}  \\  
		& H-CNN      & 99.45\%  & 99.4\%  & 99.2\%  & 99.2\%     \\   \hline
		
		\multirow{4}{*}{70}    &De Rosa[9]    & 68.68\%  & 65.61\%  & 62.24\%  & 63.93\%      \\  
		& Cao[4]    & \------  & \------    & \------ & \------  \\ 
		& Li[5]   & \textbf{99.9\%}  & \textbf{99.9\%}  & \textbf{99.9\%}  & \textbf{99.9\%}    \\   
		& P-CNN      & 99.8\%  & 99.75\%  & 99.55\%  & 99.8\%  \\  
		& H-CNN      & 97.35\%  & 98.35\%  & 97.8\%  & 98.15\%     \\   \hline 
	\end{tabular}
\end{table}

\subsection{Effect of the scale of training data}
It is well known that the scale of data has an important effect on performance for deep-learning based method. In this part, we conducted experiments to evaluate the effect of the scale of data on performance of H-CNN and P-CNN. The images from BOSSBase are firstly cropped into 128x128 pixel patches with non-overlapping. Then these images are enhanced with $\gamma=0.6$. We randomly chose 80000 image pairs as test data and 5000, 20000, 40000, 80000 image pairs as training datas. Four groups of H-CNN, P-CNN are generated using above four training datas and the test data is same for these experiments. The result is as shown in Figure.4. It can be seen that the scale of training data has an slight effect on H-CNN with small parameters and the opposite happens for P-CNN. Therefore, larger scale of training data be beneficial to the performance of P-CNN and the performance of P-CNN would be improved by increasing training data.

\begin{figure}[h]
	\centering
	\includegraphics[width=6cm,height=4cm]{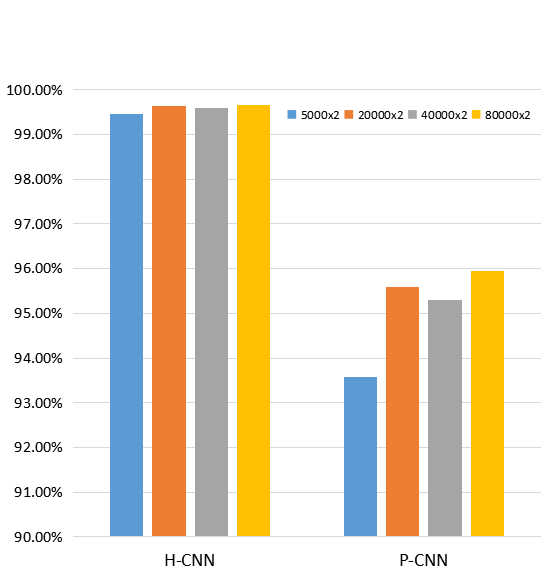}
	\caption{Effect of the scale of training data.}
\end{figure}

\section{Conclusion}
The existing schemes for contrast enhancement forensics have an unsatisfactory performance, especially, in the cases of pre-JPEG comression and anti-forensic attacks. To deal with such problems, in this paper two robust CE forensics algorithms based on deep learning (H-CNN, P-CNN) are proposed. Such methods achieve end-to-end classification based on pixel and histogram domain. Experimental results show that our proposed H-CNN attains better performance than the state-of-the-art ones in the case of no other operation and proposed methods are robust against pre-JPEG compression, anti-forensic attack, and CE level variation.

\end{document}